\documentclass[prb,twocolumn,showpacs]{revtex4}
\usepackage{color,soul}
\usepackage{amsmath}
\usepackage{dcolumn}
\usepackage{graphicx}
\usepackage{bm}
\usepackage{amssymb}
\usepackage{subfigure}
\begin{document}

\title{The optical conductivity of the 2D $t-J$ model and the origin of electron incoherence in the high-T$_{c}$ cuprate superconductors: a variational study}
\author{Jianhua Yang and Tao Li}
\affiliation{Department of Physics, Renmin University of China, Beijing 100872, P.R.China}

\begin{abstract}
Understanding the origin of electron incoherence is the first step toward a theoretical description of the non-Fermi liquid behavior of the high-T$_{c}$ cuprate superconductors. Such electron incoherence manifests itself most evidently in the non-Drude behavior of the optical response of the system and the anomalous density fluctuation behavior in the long wave length limit. The spectral weight transfer related to such dissipative response, which is absent in conventional Fermi liquid metal, has direct consequence on the dc transport property of the system in the normal state and the superfluid stiffness in the superconducting state. It is found that such electron incoherence remains significant even in the clean limit and at low temperature and thus must be attributed to the strong electron correlation effect in the cuprate superconductors. Here we study such an intrinsic effect in the 2D $t-J$ model through the variational calculation of its optical conductivity $\sigma(\omega)$. We assume a resonating valence bond ground state as our starting point and find that a significant portion of the total optical spectral weight remains incoherent throughout the phase diagram. The optical absorption is found to extend all the way to an energy of the order of the bare band width. We find that both the total optical weight $\bar{K}$ and the integrated incoherent optical weight $I$ increase monotonically with doping, with their ratio $R_{incoh}=I/\bar{K}$ decreasing monotonically with doping. Our results indicate that the majority part of electron incoherence in the 2D $t-J$ model can be attributed to the electron fractionalization mechanism assumed in such a treatment. We also find that the Drude weight deduced from $D=\bar{K}-I$ scales linearly with hole doping, without any sign of a non-monotonic behavior in the overdoped regime. Our results form an estimate of the lower bound for electron incoherence in the 2D $t-J$ model as the multi-spinon excitation processes are neglected in our treatment.   
\end{abstract}

\maketitle

\section{Introduction}
A systematic understanding of the non-Fermi liquid behavior is believed to be the key step to resolve the mystery of the high T$_{c}$ superconductivity in the cuprate superconductors. The most well known example of such non-Fermi liquid behavior in the cuprate superconductors is the perfect linear-in-T dc resistivity from very high temperature all the way down to extremely low temperature\cite{Cooper,Taillefer,Hussey,Proust,Phillips}. While it is generally believed that the strong electron correlation effect in the cuprate superconductors is essential in the origin of such a non-Fermi liquid behavior, other more extrinsic mechanisms such as the electron-phonon coupling and disorder effect, together with their interplay with the electron correlation effect, may also be indispensable to fully account for such a non-Fermi liquid behavior in the dc limit\cite{Li1}. This greatly complicates the theoretical analysis of such non-Fermi liquid behaviors.  

At a more microscopic time scale, the non-Fermi liquid behavior also manifests itself in various electron spectrums. For example, even at the early stage of the high-T$_{c}$ era people had already noticed that the electronic Raman spectrum of the cuprate superconductors is anomalous in that it features an extremely broad continuum extending to the energy scale of the band width\cite{Bosovic,Varma}. This is very different from what one would expect in a conventional Fermi liquid metal, in which the generalized density fluctuation in the electronic Raman excitation should be restricted to an energy window of the order of $v_{F}q$. Here $v_{F}$ and $q$ are the Fermi velocity and the wave vector of the photon respectively. More recently, anomalous density fluctuation has been directly inferred from the electron energy loss spectrum(EELS) measurement on the cuprate superconductors\cite{Mitrano}, in which a momentum and energy independent broad continuum(when normalized by $q^{2}$) has been reported. This is totally different from what we would expect from the particle-hole excitation around the Fermi surface in a conventional Fermi liquid. Beside the density fluctuation, the non-Fermi liquid behavior also manifests itself clearly in the optical absorption spectrum of the cuprate superconductors\cite{Marel,Heumen}, which is found to extend all the way to the energy scale of the band width. The optical conductivity of the cuprate superconductors is found to exhibit a very slow decay in frequency of the form $\omega^{-\alpha}$, with $\alpha$ generally an irrational number less than one. Such an anomalous behavior, which is usually called the non-Drude behavior in the literature, signifies the importance of electron incoherence in the optical excitation process. 

Unlike the non-Fermi liquid behavior in the dc limit, the non-Fermi liquid behavior in the optical response and the density fluctuation spectrum is believed to be more intrinsic, since neither the electron-phonon coupling nor the impurity scattering is expected to contribute to electron incoherence at the energy scale comparable to the band width. It is thus hopeful that such non-Fermi liquid behaviors can be understood from model study of the cuprate superconductors involving only electron degree of freedom, for example, the study of the 2D $t-J$ model. We note that a study of electron incoherence at the energy scale of the band width can also shed important light on the dc transport behavior of the system. More specifically, the integrated incoherent optical weight, denoted below as $I$, is nothing but a measure of the reduction in the spectral weight participating the dc transport from the total optical spectral weight $\bar{K}$. It thus determines directly the dc resistivity in the normal state. In the superconducting state, $I$ measures the reduction of the superfluid density from $\bar{K}$. The transfer of spectral weight between the high energy and the low energy regime is believed to be a major mechanism for the emergence of the non-Fermi liquid behavior in doped Mott insulators\cite{Phillips}.   

In a purely electronic model, electron incoherence is conventionally attributed to the scattering with some collective modes in the system, for example, the thermal or quantum fluctuation in the spin or the charge channel. However, such collective mode usually gain appreciate spectral weight only when the system is close to the instability toward some symmetry breaking phase and thus has a small energy scale. It is thus unlikely that the scattering with such collective modes to be responsible for the electron incoherence at the energy scale of the band width. To generate electron incoherence at such a high energy scale, a broad spectral continuum is needed. Such continuum should also be ubiquitous in the phase diagram of the cuprate superconductors, since the non-Fermi liquid behavior is not limited to any particular doping\cite{Phillips}.

The fluctuation of the local moment in a doped Mott insulator may just provide such a broad and ubiquitous continuum. In a doped Mott insulator such as the cuprate superconductor, the local moment remains well defined even when the magnetic long range order is eliminated, since its robustness is protected by large energy scale, for example, the Hubbard interaction $U$ in the Hubbard model. Indeed, RIXS measurements in the last decade\cite{Tacon,Dean} find that spin-wave-like paramagnon fluctuation exists ubiquitously in the phase digram of the cuprate superconductors, with its energy scale and integrated intensity almost doping independent. This is consistent with what we would expect for the fluctuation of local moment in a quantum paramagnetic state, or, a quantum spin liquid state. It is very likely that the electron incoherence as manifested in the non-Drude optical absorption behavior has its origin in such ubiquitous paramagnetic fluctuation. Evidence for the strong coupling of the electron with spectral weight of such a spectral character has indeed be found in time resolved optical measurement\cite{Conte}.

A theoretical description of the local moment fluctuation in a quantum spin liquid state is already a big theoretical challenge, not to say its possible relevance to the origin of electron incoherence. Here we try to answer this question from the perspective of the resonating valence bond(RVB) theory of the 2D $t-J$ model\cite{Lee}. Within the RVB theory, the doped cuprate superconductor is described by a doped quantum spin liquid, in which the electron fractionalizes into charge neutral spinon and spinless holon. The spinon is assumed to carry the spin quantum number of an electron and describes the fluctuation of the local moment in a quantum spin liquid. The holon is assumed to carry the charge quantum number of an electron and describes the charge backflow accompanying spinon motion. Such charge backflow is necessary as a result of the no double occupancy constraint on the electron operator in the $t-J$ model. The very fractionalization of electron provides a new mechanism of electron incoherence other than the scattering of electrons. One advantage of the spin-charge separation mechanism is that it naturally results in electron incoherence at the energy scale of the electron band width.  

However, a quantitative analysis of the spin-charge separation effect on the electron incoherence of the cuprate superconductor is difficult. As we will see in the following, a self-consistent RVB mean field theory treatment of the $t-J$ model\cite{Kotliar} results in the unphysical prediction of a negative Drude weight. To cure such a problem, we have calculated the optical conductivity of the 2D $t-J$ model using a dynamical variational theory. The theory calculate the dynamical behavior of a strongly correlated model by diagonalizing the Hamiltonian within a variational subspace spanned by a chosen set of strongly correlated basis functions\cite{Li2}. For the 2D $t-J$ model studied here, the basis function is generated by Gutzwiller projection of mean field excitation above a RVB ground state. This requires the treatment of a large number of strongly correlated basis function in the computation. We have improved the algorithm to speed up such a calculation.

Our calculation indicates that the spin-charge separation mechanism is responsible for a majority part of electron incoherence in the 2D $t-J$ model. More specifically, we find that both the total optical weight $\bar{K}$ and the integrated incoherent optical weight $I$ increase monotonically with doping, with their ratio $R_{incoh}=I/\bar{K}$ decreasing monotonically with doping. More than half of total optical weight remains incoherent throughout the phase diagram. We also find that the incoherent spectral weight constitutes a very broad continuum extending to energy as high as the band width. The Drude weight deduced from $D=\bar{K}-I$ scales linearly with hole doping, without any sign of a non-monotonic behavior. This implies that the decrease of the superfluid stiffness observed in the overdoped regime may have some extrinsic origin\cite{Bozovic1,Michon}.

The paper is organized as follows. In the next section, we introduce the 2D $t-J$ model studied in this paper and the general formula for its optical conductivity. We then present a self-consistent RVB mean field theory of the 2D $t-J$ model in the third section and show that the mean field treatment fails to provide a consistent description of the electron incoherence in this model. In the fourth section, we present the dynamical variational theory for the optical conductivity of the 2D $t-J$ model. We will also introduce the algorithm improvement which makes possible the real computation of these quantities. In the last section, we discuss the implication of our result and the possible extensions of it.

\section{The optical conductivity of the 2D $t-J$ model}
The 2D $t-J$ model studied in this work is given by the following Hamiltonian 
\begin{eqnarray}
H=&-&t\sum_{i,\bm{\delta},\alpha}(c^{\dagger}_{i,\alpha}c_{i+\bm{\delta},\alpha}+h.c.)\nonumber\\
&-&t'\sum_{i,\bm{\delta}',\alpha}( c^{\dagger}_{i,\alpha}c_{i+\bm{\delta}',\alpha}+h.c.)\nonumber\\
&+&J\sum_{i,\bm{\delta}}(\mathbf{S}_{i}\cdot \mathbf{S}_{i+\bm{\delta}}-\frac{1}{4}n_{i}n_{i+\bm{\delta}}),
\end{eqnarray}
Here $\alpha=\uparrow,\downarrow$ denotes the spin of the electron. $t$ and $t'$ denote the hopping integral between the nearest-neighboring and the next-nearest-neighboring sites of the square lattice. $J$ is the Heisenberg exchange coupling between nearest-neighboring spins. $\bm{\delta}=\bm{x,y}$ denotes the nearest-neighboring vector on the square lattice. $\bm{\delta}'=\bm{x\pm y}$ denotes the next-nearest-neighboring vector on the square lattice. The electron operator $c^{\dagger}_{i,\sigma}$ satisfies the following constraint of no double occupancy
\begin{equation}
\sum_{\sigma}c^{\dagger}_{i,\sigma}c_{i,\sigma}\leq 1
\end{equation}
$\mathbf{S}_{i}$ and $n_{i}$ are the spin and electron number operator on site $i$. In this study, we choose $t'=-0.3t$, $J=t/3$. We will use $t$ as the unit of energy.

To calculate the optical conductivity of the system, we couple the electron to a spatial uniform electromagnetic potential $\mathbf{A}(t)$ through the Peierls substitution. We assume that $\mathbf{A}(t)$ is directed along the $x$-direction of the square lattice, namely, $\mathbf{A}(t)=(A(t),0)$. The Hamiltonian then reads 
\begin{eqnarray}
H[\ \mathbf{A}(t)\ ]=&-&t\sum_{i,\bm{\delta},\alpha}(e^{i\mathbf{A}(t)\cdot \bm{\delta}}c^{\dagger}_{i,\alpha}c_{i+\bm{\delta},\alpha}+h.c.)\nonumber\\
&-&t'\sum_{i,\bm{\delta}',\alpha}( e^{i\mathbf{A}(t)\cdot\bm{\delta}'}c^{\dagger}_{i,\alpha}c_{i+\bm{\delta}',\alpha}+h.c.)\nonumber\\
&+&J\sum_{i,\bm{\delta}}(\mathbf{S}_{i}\cdot \mathbf{S}_{i+\bm{\delta}}-\frac{1}{4}n_{i}n_{i+\bm{\delta}}),
\end{eqnarray}
Here we have adopted the convention $\hbar=e=c=a=1$ for convenience, in which $a$ denotes the lattice constant of the square lattice. Here we note that $\mathbf{A}(t)\cdot \bm{x}=\mathbf{A}(t)\cdot \bm{\delta'}=A(t)$, $\mathbf{A}(t)\cdot \bm{y}=0$.

Expanding the Hamiltonian to the second order in $A(t)$, we have 
\begin{equation}
H[\ \mathbf{A}(t)\ ]\approx H-j^{x}_{p}A(t)+\frac{1}{2}KA^{2}(t) 
\end{equation}
in which 
\begin{eqnarray}
j^{x}_{p}=&it&\sum_{i,\alpha}(c^{\dagger}_{i,\alpha}c_{i+\bm{x},\alpha}-h.c.)\nonumber\\
&+it'&\sum_{i,\bm{\delta}',\alpha}(c^{\dagger}_{i,\alpha}c_{i+\bm{\delta}',\alpha}-h.c.),
\end{eqnarray}
is the paramagnetic current along the $x$-direction. 
\begin{eqnarray}
K=&t&\sum_{i,\alpha}(c^{\dagger}_{i,\alpha}c_{i+\bm{x},\alpha}+h.c.)\nonumber\\
&+t'&\sum_{i,\bm{\delta}',\alpha}(c^{\dagger}_{i,\alpha}c_{i+\bm{\delta}',\alpha}+h.c.),
\end{eqnarray}
is the kernel of diamagnetic response along the $x$-direction.

Using the Kubo formula, the optical conductivity of the model is given by
\begin{equation}
\Re \sigma(\omega)=D\delta(\omega)+\sigma^{reg}(\omega)
\end{equation}
in which the regular part of the optical conductivity is given by
\begin{equation}
\sigma^{reg}(\omega)=-\frac{\Im \Pi(\omega+i0^{+})}{\omega}
\end{equation}
Here $ \Pi(\omega+i0^{+})$ is the Fourier transform of the retarded green function of the current operator. It is defined by
\begin{equation}
\Pi(t)=-i\theta(t)\langle [j^{x}_{p}(t),j^{x}_{p}(0) ]  \rangle
\end{equation}
in which $j^{x}_{p}(t)$ is the Heisenberg operator of $j^{x}_{p}$. The average is done in the ground state of the model. The Drude weight $D$ is given by 
\begin{equation}
D=  \bar{K}  + \Re  \Pi (i0^{+})
\end{equation}
Here $\bar{K}=\langle K \rangle$ denotes the ground state expectation value of $K$ and is a measure of the total optical spectral weight along the $x$-direction. Using the Kronig-Kramers relation between $\Im \Pi(\omega+i0^{+})$ and $\Re \Pi(\omega+i0^{+})$, we have
\begin{equation}
D=\bar{K}-2\int_{0^{+}}^{\infty} d\omega \sigma^{reg}(\omega)
\end{equation}
The meaning of this equation is self-evident: the Drude weight is given by the difference between the total optical spectral weight in the $x$-direction and the integrated spectral weight involved in optical absorption. In the superconducting state, the Drude weight is just the superfluid density of the system.

More specifically, we can define the fraction of the incoherent spectral weight $I$ as follows
\begin{equation}
I=2\int_{0^{+}}^{\infty} d\omega \sigma^{reg}(\omega)
\end{equation}
Using Eq.8 and the relation 
\begin{equation}
-2\Im \Pi(\omega+i0^{+})=2\pi\sum_{n}|\langle n| j^{x}_{p} |g\rangle|^{2}\delta[\omega-(E_{n}-E_{g})]
\end{equation}
we have
\begin{equation}
I=2\pi\sum_{n}\frac{|\langle n| j^{x}_{p} |g\rangle|^{2}}{E_{n}-E_{g}}
\end{equation}
Here $| g \rangle$ denotes the ground state of the system. $E_{g}$ is the ground state energy, $| n \rangle$ is the excited state with energy $E_{n}$.  
From $I$ we can define the incoherence ratio as follows
\begin{equation}
R_{incoh}=\frac{I}{\bar{K} }
\end{equation}
It measures how much electron spectral weight is involved in the optical absorption process.

In a free electron system, $j^{x}_{p}$ is a conserved quantity and is identically zero in the ground state. As a result, the electron is entirely coherent and $R_{incoh}=0$. There is thus no optical absorption at any nonzero frequency and $D=\bar{ K}$. For a translational invariant electron model, any electron incoherence should thus be attributed to interaction effect. In the $t-J$ model studied here, $j^{x}_{p}$ is no longer a conserved quantity for two reasons. First, the electron operator is now subjected to the no double occupancy constraint and is not a free fermion operator. Second, there is the scattering related to the Heisenberg exchange coupling. In this work, we will treats both effect in the fermionic RVB framework.

\section{The failure of a RVB mean field theory for the optical conductivity of the 2D $t-J$ model}
In the RVB framework, we rewrite the electron operator in terms of the spinon and the holon operator as follows
\begin{equation}
c^{\dagger}_{i,\alpha}=f^{\dagger}_{i,\alpha}b_{i}
\end{equation}  
Here $f^{\dagger}_{i,\alpha}$ is the charge neutral fermionic spinon operator with spin $\alpha$ on site $i$. $b_{i}$ is the spinless holon operator carrying the charge. Such a representation is exact if the following constraint is satisfied
\begin{equation}
\sum_{\alpha}f^{\dagger}_{i,\alpha}f_{i,\alpha}+b^{\dagger}_{i}b_{i}=1
\end{equation}
In terms of the spinon and the holon operators, the $t-J$ Hamiltonian now reads, 
\begin{eqnarray}
H=&-&t\sum_{i,\bm{\delta},\alpha}(f^{\dagger}_{i,\alpha}f_{i+\bm{\delta},\alpha}b^{\dagger}_{i}b_{i+\bm{\delta}}+h.c.)\nonumber\\
&-&t'\sum_{i,\bm{\delta}',\alpha}( f^{\dagger}_{i,\alpha}f_{i+\bm{\delta}',\alpha}b^{\dagger}_{i}b_{i+\bm{\delta}'}+h.c.)\nonumber\\
&+&\frac{J}{4}\sum_{i,\bm{\delta},\alpha,\beta,\gamma,\eta}(f^{\dagger}_{i,\alpha}\bm{\sigma}_{\alpha,\beta}f_{i,\beta})\cdot (f^{\dagger}_{i+\bm{\delta},\gamma}\bm{\sigma}_{\gamma,\eta}f_{i+\bm{\delta},\eta})\nonumber\\
&-&\frac{J}{4}\sum_{i,\bm{\delta},\alpha,\beta}f^{\dagger}_{i,\alpha}f_{i,\alpha}f^{\dagger}_{i+\bm{\delta},\beta}f_{i+\bm{\delta},\beta},
\end{eqnarray} 
in which $\alpha,\beta,\gamma,\eta=\uparrow,\downarrow$. Here we have represented the spin operator as 
\begin{equation}
\mathbf{S}_{i}=\frac{1}{2}\sum_{\alpha,\beta}f^{\dagger}_{i,\alpha}\bm{\sigma}_{\alpha,\beta}f_{i,\beta}
\end{equation}
in which $\bm{\sigma}$ is the usual Pauli matrix for electron spin. 

In the RVB mean field theory, which is also called the slave Boson mean field theory, we introduce the following RVB order parameters for the spinon 
\begin{eqnarray}
\chi&=&\langle \sum_{\alpha}f^{\dagger}_{i,\alpha}f_{i+\bm{x},\alpha} \rangle=\langle \sum_{\alpha}f^{\dagger}_{i,\alpha}f_{i+\bm{y},\alpha} \rangle\nonumber\\
\chi'&=&\langle \sum_{\alpha}f^{\dagger}_{i,\alpha}f_{i+\bm{x+y},\alpha} \rangle=\langle \sum_{\alpha}f^{\dagger}_{i,\alpha}f_{i+\bm{x-y},\alpha} \rangle\nonumber\\
\Delta&=&\langle \sum_{\alpha}s_{\alpha} f^{\dagger}_{i,\alpha}f^{\dagger}_{i+\bm{x},\bar{\alpha}} \rangle=-\langle \sum_{\alpha}s_{\alpha} f^{\dagger}_{i,\alpha}f^{\dagger}_{i+\bm{y},\bar{\alpha}} \rangle
\end{eqnarray}
in which $\bar{\alpha}$ denotes the inverse of $\alpha$, $s_{\uparrow}=1$, $s_{\downarrow}=-1$. Here we have assumed that all these RVB order parameters are translational invariant. We also assume that the holon all condense into the $\mathbf{q}=0$ state at zero temperature. Thus we have
\begin{equation}
\langle b_{i} \rangle=\langle b^{\dagger}_{i} \rangle=\sqrt{x}
\end{equation}
Here $x$ denotes the hole density in the CuO$_{2}$ plane.

Decoupling the $t-J$ Hamiltonian with the above order parameters leads to the following RVB mean field Hamiltonian for both the spinon and the holon
\begin{eqnarray}
H_{f}&=&\sum_{\mathbf{k},\alpha}\epsilon^{f}_{\mathbf{k}}\  f^{\dagger}_{\mathbf{k},\alpha}f_{\mathbf{k},\alpha}-\sum_{\mathbf{k}}\Delta_{\mathbf{k}}(f^{\dagger}_{\mathbf{k},\uparrow}f^{\dagger}_{-\mathbf{k},\downarrow}+f_{-\mathbf{k},\downarrow}f_{\mathbf{k},\uparrow})\nonumber\\
H_{b}&=&\sum_{\mathbf{q}}\epsilon^{b}_{\mathbf{q}}\  b^{\dagger}_{\mathbf{q}}b_{\mathbf{q}}
\end{eqnarray}
with
\begin{eqnarray}
\epsilon^{f}_{\mathbf{k}}=&-&2(tx+\frac{3J\chi}{8})(\cos k_{x}+\cos k_{y})\nonumber\\
&-&4t'x\cos k_{x} \cos k_{y}-\mu_{f}\nonumber\\
\Delta_{\mathbf{k}}=&&\frac{3J\Delta}{4} (\cos k_{x}-\cos k_{y})\nonumber\\
\epsilon^{b}_{\mathbf{q}}=&-&2t\chi(\cos q_{x}+\cos q_{y})\nonumber\\
&-&4t'\chi' \cos q_{x} \cos q_{y}-\mu_{b}
\end{eqnarray}
Here $\mu_{f}$ denotes the chemical potential of the spinon. It is determined by the following spinon density equation
\begin{equation}
\langle \sum_{\alpha} f^{\dagger}_{i,\alpha}f_{i,\alpha}\rangle=1-x
\end{equation}
$\mu_{b}$ is the chemical potential of the holon. It is determined by the requirement of $\epsilon^{b}_{\mathbf{q}=0}=0$. Thus in the RVB mean field theory the no double occupancy constraint Eq.17 is relaxed to a constraint in the average sense. We will see in the following that such a relaxation of the no double constraint will result in inconsistency in the optical conductivity of the system.

The spinon Hamiltonian $H_{f}$ can be diagonalized by the Bogliubov transformation of the form
\begin{equation}
\left(\begin{array}{c} f_{\mathbf{k},\uparrow} \\f^{\dagger}_{-\mathbf{k},\downarrow} \end{array}\right)=\left(\begin{array}{cc}u_{\mathbf{k}} & v_{\mathbf{k}} \\-v_{\mathbf{k}} & u_{\mathbf{k}}\end{array}\right)\left(\begin{array}{c} \gamma_{\mathbf{k},\uparrow} \\\gamma^{\dagger}_{-\mathbf{k},\downarrow} \end{array}\right)
\end{equation}
$\gamma_{\mathbf{k},\alpha}$ is the quasiparticle operator with energy 
\begin{equation}
E_{\mathbf{k}}=\sqrt{(\epsilon^{f}_{\mathbf{k}})^{2}+(\Delta_{\mathbf{k}})^{2}}
\end{equation}
We then solve the self-consistent equation for the RVB parameters $\chi$, $\chi'$ and $\Delta$ from their defining equation Eq.20 in the mean field ground state $|\mathrm{MF}\rangle$, which takes the form of
\begin{equation}
|\mathrm{MF}\rangle=\frac{1}{\sqrt{N_{b}!}}(b^{\dagger}_{\mathbf{q}=0})^{N_{b}}|f-\mathrm{BCS}\rangle
\end{equation}
Here $N_{b}=xN$ is the number of doped holes, $N$ is the number of lattice site. $|f-\mathrm{BCS}\rangle$ denotes the ground state of spinon Hamiltonian $H_{f}$. The details of the above derivations can be found in Ref.[\onlinecite{Kotliar}]. In Fig.1, we present the mean field phase diagram obtained from the solution of the self-consistent equations Eq.20 and Eq.24.

\begin{figure}
\includegraphics[width=7.5cm]{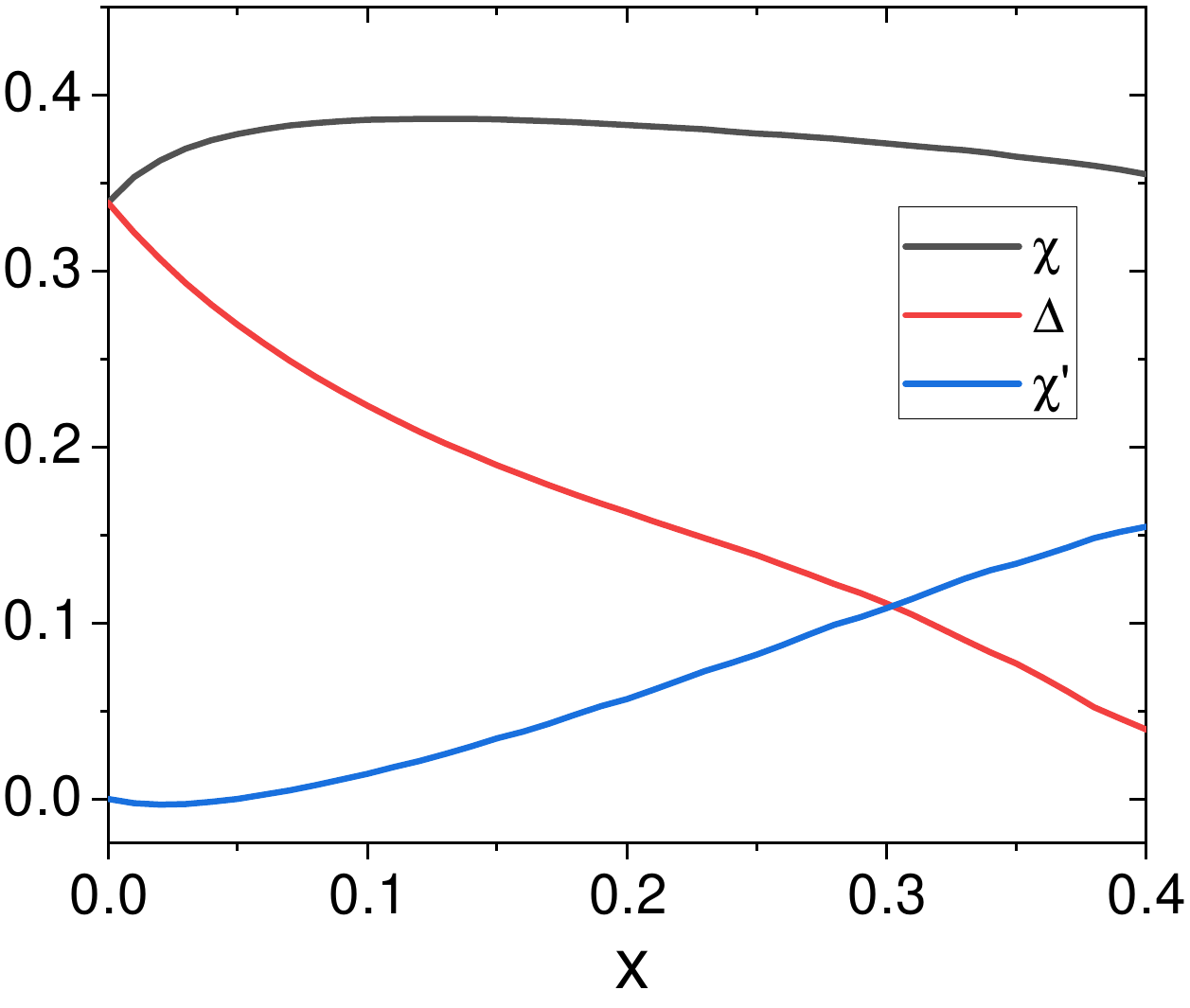}
\caption{The mean field phase digram of the 2D $t-J$ model obtained from the solution of the self-consistent equations Eq.20 and Eq.24 in the RVB mean field ground state $|\mathrm{MF}\rangle$.}
\end{figure}

We now turn to the calculation of the optical conductivity above the RVB mean field ground state $|\mathrm{MF}\rangle$. In terms of the spinon and the holon operators, the paramagnetic current operator $j^{x}_{p}$ can be rewritten as
\begin{equation}
j^{x}_{p}=\frac{1}{N}\sum_{\mathbf{k,k',q},\alpha} v^{x}_{\mathbf{q-k}} f^{\dagger}_{\mathbf{k},\alpha}f_{\mathbf{k}',\alpha}b^{\dagger}_{\mathbf{q+k'-k}}b_{\mathbf{q}}
\end{equation} 
in which
\begin{equation}
v^{x}_{\mathbf{k}}=2t \sin k_{x}+4t'\sin k_{x}\cos k_{y}
\end{equation}
is the $x$-component of the band velocity. We now apply $j^{x}_{p}$ on the mean field ground state $|\mathrm{MF}\rangle$. After some algebraic manipulation, we arrive at
\begin{eqnarray}
j^{x}_{p}|\mathrm{MF}\rangle&=&\sum_{\mathbf{q\neq0,k}}  \Phi_{\mathbf{k,q}}\ \gamma^{\dagger}_{\mathbf{k},\uparrow}\gamma^{\dagger}_{-(\mathbf{k+q}),\downarrow}\nonumber\\
&\times&\frac{1}{\sqrt{(N_{b}-1)!}}b^{\dagger}_{\mathbf{q}}(b^{\dagger}_{\mathbf{q}=0})^{N_{b}-1}|f-\mathrm{BCS}\rangle
\end{eqnarray} 
in which 
\begin{equation}
\Phi_{\mathbf{k,q}}=\frac{\sqrt{N_{b}}}{N}[v^{x}_{\mathbf{k}}u_{\mathbf{k}}v_{\mathbf{k+q}}-v^{x}_{\mathbf{k+q}}v_{\mathbf{k}}u_{\mathbf{k+q}}]
\end{equation}
For later convenience, we define
\begin{equation}
|\mathbf{k,q}\rangle=\gamma^{\dagger}_{\mathbf{k},\uparrow}\gamma^{\dagger}_{-(\mathbf{k+q}),\downarrow}\frac{1}{\sqrt{(N_{b}-1)!}}b^{\dagger}_{\mathbf{q}}(b^{\dagger}_{\mathbf{q}=0})^{N_{b}-1}|f-\mathrm{BCS}\rangle
\end{equation}
we then have
\begin{equation}
j^{x}_{p}|\mathrm{MF}\rangle=\sum_{\mathbf{q\neq0,k}} \Phi_{\mathbf{k,q}}|\mathbf{k,q}\rangle
\end{equation}
We note that $|\mathbf{k,q}\rangle$ is a normalized mean field excited state with energy $\epsilon^{b}_{\mathbf{q}}+E_{\mathbf{k}}+E_{\mathbf{k+q}}$. 

The optical conductivity is then given by
\begin{eqnarray}
&\sigma^{reg}(\omega)&=-\frac{\Im \Pi(\omega+i0^{+})}{\omega}\nonumber\\
&=\pi&\sum_{\mathbf{q\neq0,k}}\frac{|\Phi_{\mathbf{k,q}}|^{2}}{\epsilon^{b}_{\mathbf{q}}+E_{\mathbf{k}}+E_{\mathbf{k+q}}}\delta[\omega-(\epsilon^{b}_{\mathbf{q}}+E_{\mathbf{k}}+E_{\mathbf{k+q}})]\nonumber\\
\end{eqnarray}
As a result we have
\begin{equation}
I=2\pi\sum_{\mathbf{q\neq0,k}}\frac{|\Phi_{\mathbf{k,q}}|^{2}}{\epsilon^{b}_{\mathbf{q}}+E_{\mathbf{k}}+E_{\mathbf{k+q}}}
\end{equation}
On the other hand, the total optical spectral weight is given by
\begin{equation}
\bar{K}=N_{b}(2t\chi+4t'\chi')
\end{equation}
Fig.2 presents the result of the optical conductivity calculated from the RVB mean field theory at a typical doping $x=0.15$. It is found that the optical absorption extends to a rather high energy of the order of the bare band width. This can be attributed to the fractionalization of the electron into the spinon and the holon degree of freedom, each of which carry part of the momentum of an electron. 

\begin{figure}
\includegraphics[width=7.5cm]{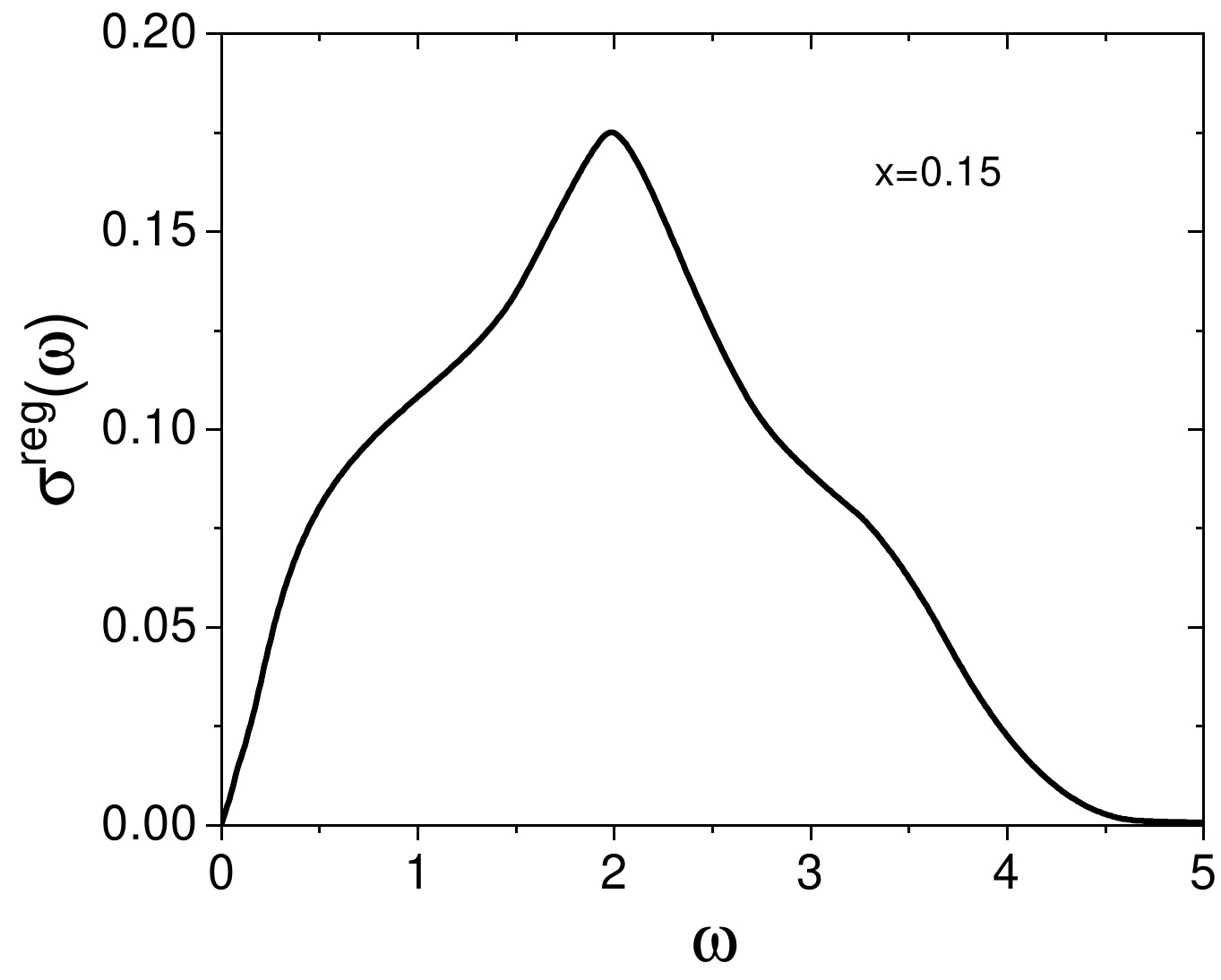}
\caption{The optical conductivity calculated from the RVB mean field theory at a typical hole doping level of $x=0.15$. $\omega$ is measured in unit of $t$.}
\end{figure}

However, a more quantitive analysis shows that there is serious internal inconsistency in the mean field treatment of the optical conductivity of the $t-J$ model. As can be seen from Fig.3, the integrated incoherent spectral weight $I$ calculated from the RVB mean field theory is much larger than the total optical weight $\bar{K}$. This would imply a negative Drude weight, a result which is obviously unphysical. On the other hand, the fractionalization of the electron should only provide an estimate of the lower bound of the electron incoherence. For example, the spinon scattering caused by the Heisenberg exchange coupling which we have ignored in the RVB mean field theory will generate additional electron incoherence. Thus the electron incoherence is greatly overestimated in the RVB mean field theory of the 2D $t-J$ model. One way to cure such a problem is to work in the projected space of no double occupancy. We will now turn to the discussion of such a variational description.

\begin{figure}
\includegraphics[width=7.5cm]{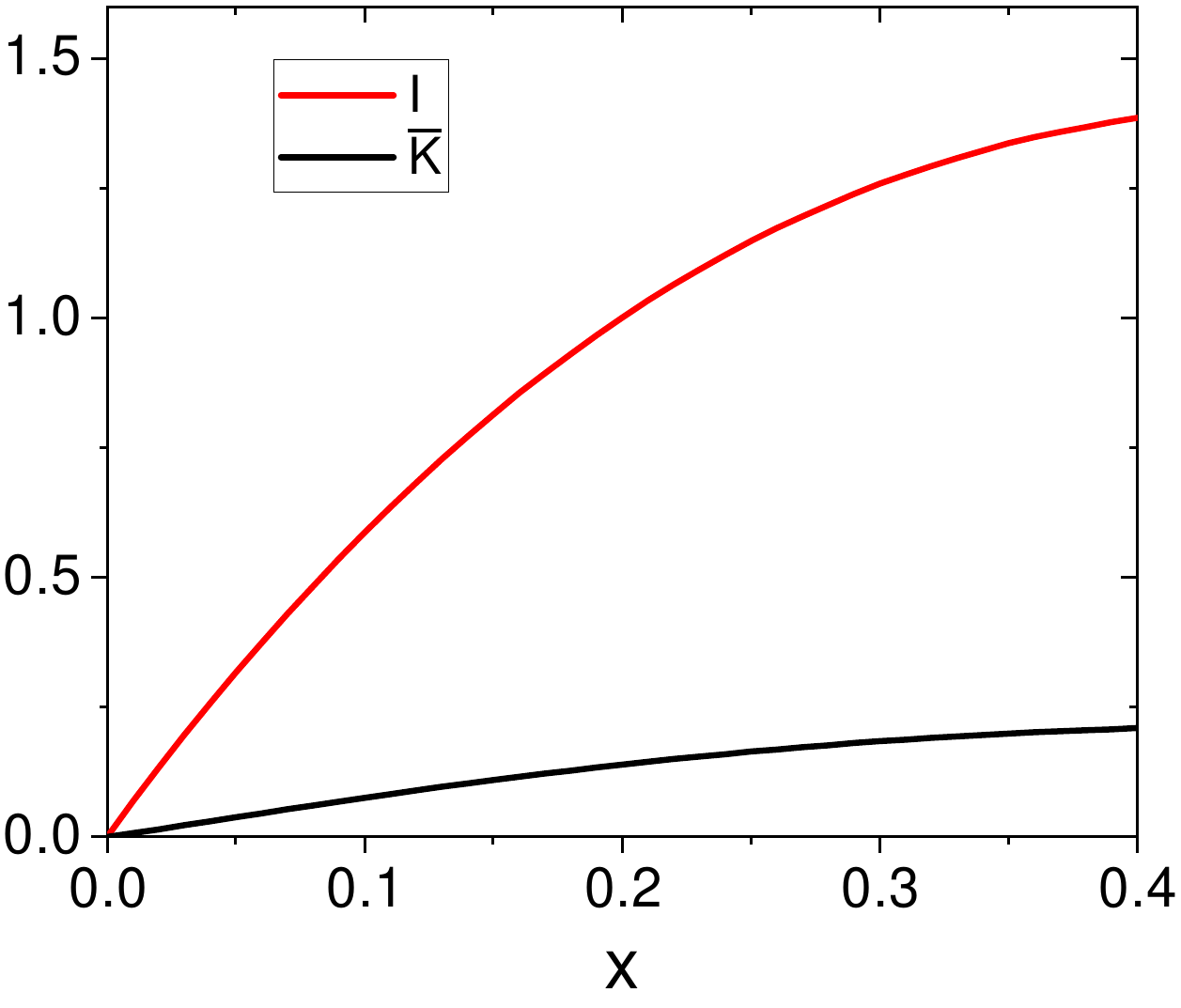}
\caption{Evolution of the total optical spectral weight $\bar{K}$ and the integrated incoherent spectral weight $I$ with $x$ calculated from the RVB mean field theory. Both $\bar{K}$ and $I$ are measured in unit of $Nt$. Note that $I$ is much larger than $\bar{K}$. This is obviously an unphysical result and it implies the internal inconsistency of the mean field treatment of the optical conductivity in the 2D $t-J$ model.}
\end{figure}

\section{A dynamical variational theory of the optical conductivity of the 2D $t-J$ model}
In the RVB theory, the ground state of the 2D $t-J$ model is given by the Gutzwiller projection of the mean field ground state, namely
\begin{equation}
|\Psi_{g}\rangle=\mathrm{P}_{\mathrm{G}}|\mathrm{MF}\rangle
\end{equation}
Here $\mathrm{P}_{\mathrm{G}}$ denotes the Gutzwiller projection that enforces the no double occupancy constraint Eq.17. The mean field ground state $|\mathrm{MF}\rangle$ is generated from a mean field ansatz of the form of $H_{f}$ with
\begin{eqnarray}
\epsilon^{f}_{\mathbf{k}}&=&-2(\cos k_{x}+\cos k_{y})-4t'_{v}\cos k_{x}\cos k_{y}-\mu_{v}\nonumber\\
\Delta_{\mathbf{k}}&=&2\Delta_{v}(\cos k_{x}-\cos k_{y})
\end{eqnarray}
in which $t'_{v}$, $\mu_{v}$ and $\Delta_{v}$ are dimensionless parameters to be determined by the optimization of the variational ground state energy. Fig.4 presents the results of the optimized variational parameters as functions of $x$. The calculation is done on a $12\times12$ cluster with periodic boundary condition.

\begin{figure}
\includegraphics[width=7.5cm]{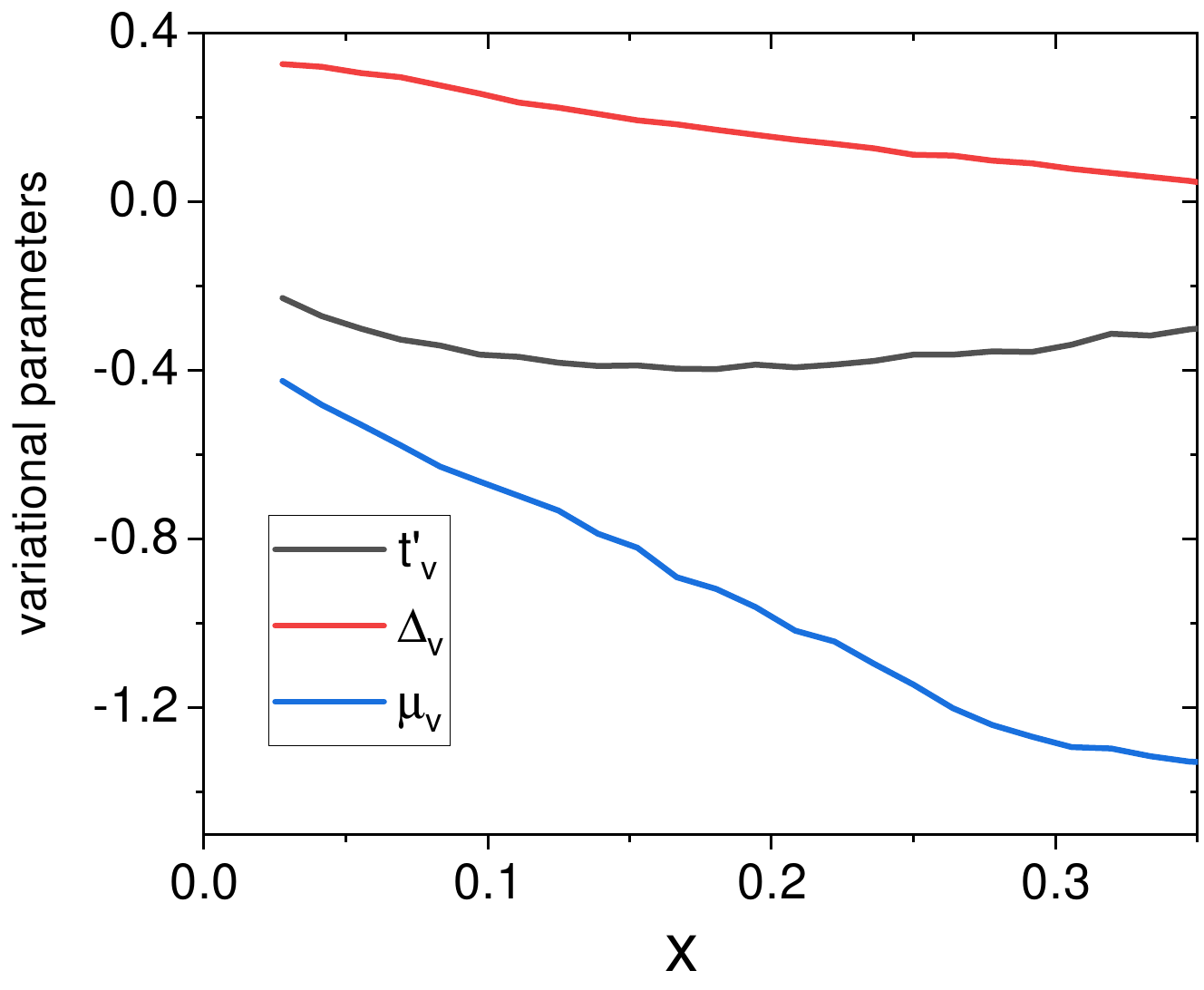}
\caption{Evolution of the optimized RVB parameters of the 2D $t-J$ model with the hole concentration $x$. The calculation is done on a $12\times12$ cluster with periodic boundary condition.}
\end{figure}

Since the variational ground state of the 2D $t-J$ model is constructed by the Gutzwiller projection of the RVB mean field ground state, it is natural to construct its variational excitations by Gutzwiller projection of the mean field excited states. This is however not quite right since the Gutzwiller projected mean field excited states are in general not orthonormal any more. In some cases the Gutzwiller projected excited states may even become linearly dependent. Nevertheless, we can interpret the Gutzwiller projected mean field excited states as the basis vectors of a variational subspace with a given quantum number. We can then re-diagonalize the Hamiltonian within such a variational subspace. This is the essence of the dynamical variational theory of the $t-J$ model proposed in Ref.[\onlinecite{Li2}]. Such a theory has been successfully applied in the study of dynamical behavior of many correlated electron systems\cite{Yang,Piazza,Ferrari,Ferrari1,Yu,Becca,Ido,Becca1,Li3,Charlebois,Penc}.

 Here we take $\overline{|\mathbf{k,q}\rangle}=\mathrm{P}_{\mathrm{G}}|\mathbf{k,q}\rangle$ as the working basis in the study of the optical conductivity of the 2D $t-J$ model. Since 
 \begin{equation}
 [\mathrm{P}_{\mathrm{G}}, j^{x}_{p}]=0
 \end{equation}
 we have
\begin{eqnarray}
j^{x}_{p}|\Psi_{g}\rangle&=&j^{x}_{p}\mathrm{P}_{\mathrm{G}}|\mathrm{MF}\rangle=\mathrm{P}_{\mathrm{G}}j^{x}_{p}|\mathrm{MF}\rangle\nonumber\\
&=&\sum_{\mathbf{k,q\neq0}}\Phi_{\mathbf{k,q}}\mathrm{P}_{\mathrm{G}}|\mathbf{k,q}\rangle\nonumber\\
&=&\sum_{\mathbf{k,q\neq0}}\Phi_{\mathbf{k,q}}\overline{|\mathbf{k,q}\rangle}
\end{eqnarray}
Here we have used Eq.33. 

In the variational subspace spanned by $\overline{|\mathbf{k,q}\rangle}$, a general eigenstate of the 2D $t-J$ model has the form
\begin{equation}
|\Psi^{(n)} \rangle=\sum_{\mathbf{k,q}} \varphi^{(n)}_{\mathbf{k,q}}\overline{|\mathbf{k,q}\rangle}
\end{equation}
in which the coefficient $\varphi^{(n)}_{\mathbf{k,q}}$ satisfies the following generalized eigenvalue equation
\begin{equation}
\sum_{\mathbf{k',q'}}\mathcal{H}_{\mathbf{k,q;k',q'}}  \varphi^{(n)}_{\mathbf{k',q'}}=E^{(n)}\sum_{\mathbf{k',q'}}\mathcal{O}_{\mathbf{k,q;k',q'}}  \varphi^{(n)}_{\mathbf{k',q'}}
\end{equation}
Here
\begin{eqnarray}
\mathcal{H}_{\mathbf{k,q;k',q'}}&=&\overline{\langle \mathbf{k,q}|} H \overline{|\mathbf{k',q'}\rangle}\nonumber\\
\mathcal{O}_{\mathbf{k,q;k',q'}}&=&\overline{\langle \mathbf{k,q}|} \ \overline{|\mathbf{k',q'}\rangle}
\end{eqnarray}
are the Hamiltonian matrix element and the overlap matrix element in the correlated basis. $E^{(n)}$ denotes the $n$-th eigenenergy of the Hamiltonian. In the following we will use $\mathcal{H}$ and $\mathcal{O}$ to denote the Hamiltonian matrix and the overlap matrix. We will also use $\varphi^{(n)}$ and $\Phi$ to denote the vector  corresponding to $|\Psi^{(n)}\rangle$ and $j_{p}^{x}|\Psi_{g}\rangle$. We note that both $\mathcal{H}$ and $\mathcal{O}$ are Hermitian matrix. In addition, $\mathcal{O}$ is positive semi-definite.

The matrix elements of $\mathcal{H}$ and $\mathcal{O}$ can be efficiently evaluated by the variational Monte Carlo(VMC) method using a highly parallelized re-weighting trick. The key to the re-weighting trick is the observation that any two correlated basis $\overline{|\mathbf{k,q}\rangle}$ differ from each other by at most a pair of spinon states and a single holon state. In practical calculation we will also perform a particle-hole transformation on the down-spin spinon. More details on the re-weighting trick and the VMC simulation of the Hamiltonian matrix and the overlap matrix can be found in Ref.[\onlinecite{Li2,Li3}].

With the eigenstate $|\Psi^{(n)} \rangle$ in hand, we can then construct the variational optical spectrum of the model. More specifically, we have
\begin{equation}
-\Im \Pi(\omega+i0^{+})=\pi\sum_{n}|\langle\Psi^{(n)} | j^{x}_{p}|\Psi_{g}\rangle|^{2}\delta[\omega-(E^{(n)}-E_{g})]
\end{equation}
Here 
\begin{equation}
E_{g}=\frac{\langle\Psi_{g}| H |\Psi_{g}\rangle}{\langle\Psi_{g}| \Psi_{g}\rangle}
\end{equation}
denotes the variational ground state energy. Using Eq.39 and Eq.40, we have
\begin{eqnarray}
\langle\Psi^{(n)} | j^{x}_{p}|\Psi_{g} \rangle&=&(\varphi^{(n)})^{\dagger}\mathcal{O}\Phi\nonumber\\
&=&\sum_{\mathbf{k,q;k',q'}}\varphi^{(n)*}_{\mathbf{k,q}} O_{\mathbf{k,q;k',q'}}\Phi_{\mathbf{k',q'}}
\end{eqnarray}
An advantage of such a dynamical variational theory is that the following ground state sum rule on the optical conductivity is automatically satisfied
\begin{eqnarray}
\int^{\infty}_{0^{+}} \omega \sigma^{reg}(\omega) d\omega&=&\int^{\infty}_{0^{+}}-\Im \Pi(\omega+i0^{+}) d\omega\nonumber\\
&=&\frac{\langle\Psi_{g}| j^{x}_{p}j^{x}_{p}|\Psi_{g}\rangle}{\langle\Psi_{g}|\Psi_{g}\rangle}
\end{eqnarray}

In practical calculation, the number of the correlated basis functions increases rapidly with the system size. For example, on the $N=12\times12$ cluster that will be adopted in the following calculation, there will be $N(N-1)=20592$ basis functions. This number can be reduced to $N(N-1)/2=10296$ if we impose the spin rotational symmetry on the basis functions, or, by working with the symmetrized basis functions of the form
\begin{equation}
\overline{|\mathbf{k,q}\rangle}_{s}=\frac{1}{\sqrt{2}}(\overline{|\mathbf{k,q}\rangle}+\overline{|\mathbf{-(k+q),q}\rangle})
\end{equation}
However, the number of such symmetrized basis functions is still prohibitively large and a full diagonalization of the generalized eigenvalue problem in this basis is still extremely expensive. Another concern with the direct diagonalization of the generalized eigenvalue problem in such a large variational subspace is the problem of numerical stability. As we have mentioned above, although the mean field excitation $|\mathbf{k,q}\rangle$ are orthonormal, the correlated basis function $\overline{|\mathbf{k,q}\rangle}$ after the Gutzwiller projection are in general not orthonormal any more. In fact, some of the correlated basis function may even become linearly dependent after the Gutzwiller projection. The overlap matrix $\mathcal{O}$ is thus in general singular.

To solve these problems, we have developed the following generalized Lanczos methods. As the first step, we regularize the overlap matrix $\mathcal{O}$ by adding to it a small positive constant matrix 
\begin{equation}
\mathcal{\tilde{O}}= \mathcal{O}+\epsilon \mathcal{I}
\end{equation}
Here $\mathcal{I}$ denotes the identity matrix and $\epsilon$ is a tiny positive number to guarantee the numerical stability of the algorithm. We then perform the Cholesky decomposition of the shifted overlap matrix $\mathcal{\tilde{O}}$
\begin{equation}
\mathcal{\tilde{O}}=L L^{\dagger}
\end{equation} 
in which $L$ is a lower triangular matrix. Defining
\begin{equation}
 \phi^{(n)}=L^{\dagger}\varphi^{(n)}
 \end{equation}
 the generalized eigenvalue problem Eq.39 can be expressed as
\begin{equation}
\tilde{\mathcal{H}}\phi^{(n)}=E^{(n)}\phi^{(n)}
\end{equation}
Here $\tilde{\mathcal{H}}=L^{-1} \mathcal{H} (L^{\dagger})^{-1}$ is a Hermitian matrix whose eigenvectors $\phi^{(n)}$ are orthonormal. Instead of solving this  standard eigenvalue problem directly, we adopted the Lanczos method to compute the spectrum of $j^{x}_{p}$. Defining 
\begin{equation}
\phi_{0}=L^{\dagger}\Phi
\end{equation}
we have 
\begin{eqnarray}
\langle\Psi^{(n)} | j^{x}_{p}|\Psi_{g} \rangle&\approx&(\varphi^{(n)})^{\dagger}\mathcal{\tilde{O}}\Phi\nonumber\\
&=&(\varphi^{(n)})^{\dagger}LL^{\dagger}\Phi=(\phi^{(n)})^{\dagger}\phi_{0}
\end{eqnarray}
Thus we have
\begin{equation}
-\Im \Pi(\omega+i0^{+})=\pi\sum_{n}|(\phi^{(n)})^{\dagger}\phi_{0}|^{2}\delta[\omega-(E^{(n)}-E_{g})]
\end{equation}
This is nothing but the spectrum of the excitation $\phi_{0}$ with respect to the Hamiltonian $\mathcal{\tilde{H}}$. It can be computed by the standard Lanczos iteration starting from the initial vector $\phi_{0}$. The detail of such a Lanczos iteration can be found in Ref.[\onlinecite{Dagotto}].

\begin{figure}
\includegraphics[width=7.5cm]{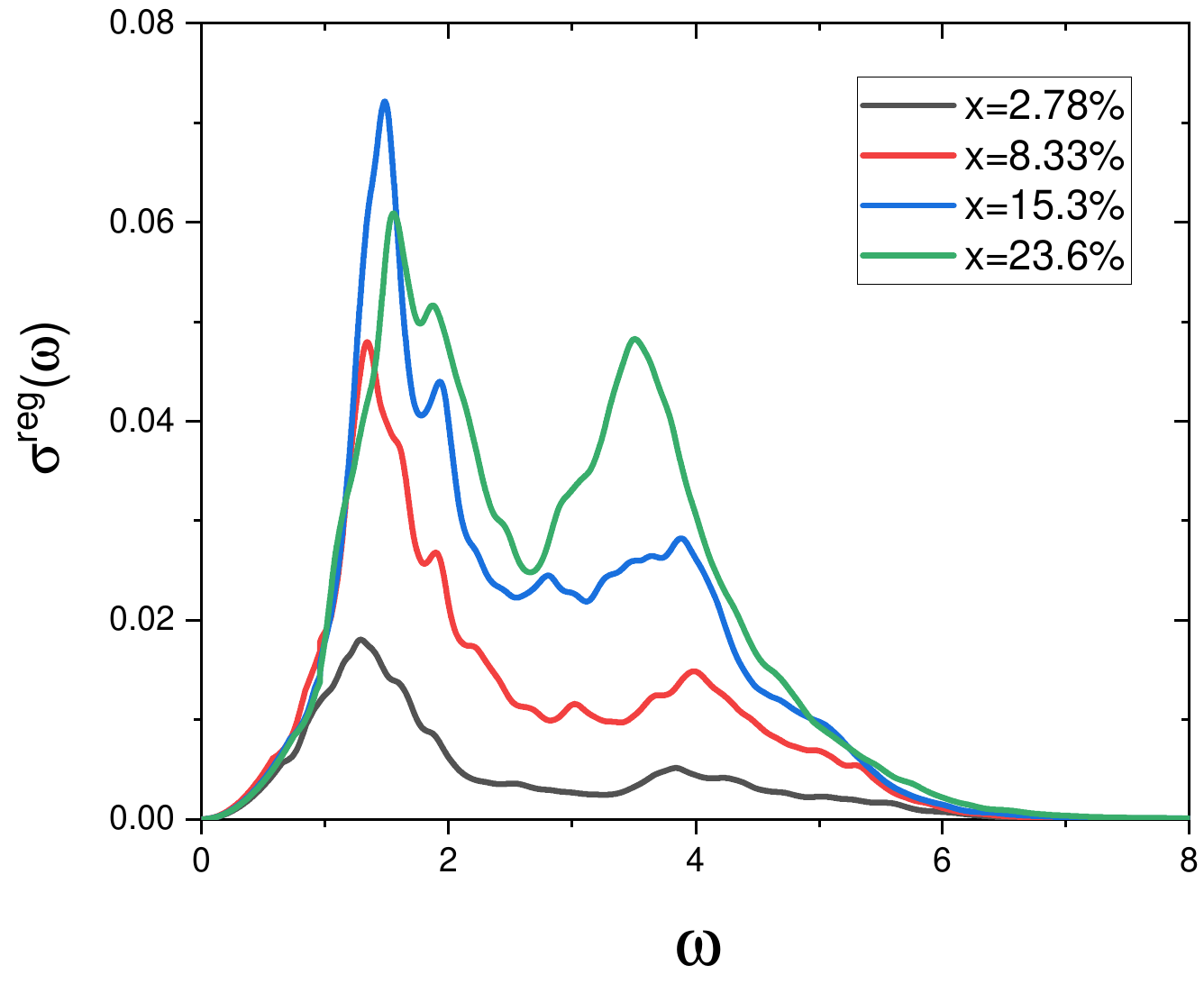}
\caption{Evolution of the optical conductivity with the hole concentration $x$ calculated from our dynamical variational theory for the 2D $t-J$ model. The calculation is done on a $12\times12$ cluster with periodic boundary condition. $\omega$ is measured in unit of $t$. Here we have neglected the antiferromagnetic ordering at low doping.}
\end{figure}

In Fig.5, we present the optical conductivity calculated from the dynamical variational theory at four different doping across the phase diagram. Since the main focus of our discussion is on the mechanism of electron incoherence in the optimal and the overdoped system, we have neglected possible antiferromagnetic long range order in the low doping regime. Similar to the mean field spectra presented in Fig.2, here the optical absorption also extends to a rather high energy of the order of the band width. The overall spectral shape of $\sigma^{reg}(\omega)$ is similar to that of the mean field spectra, with the exception that the low energy optical absorption is now strongly suppressed.

\begin{figure}
\includegraphics[width=7.5cm]{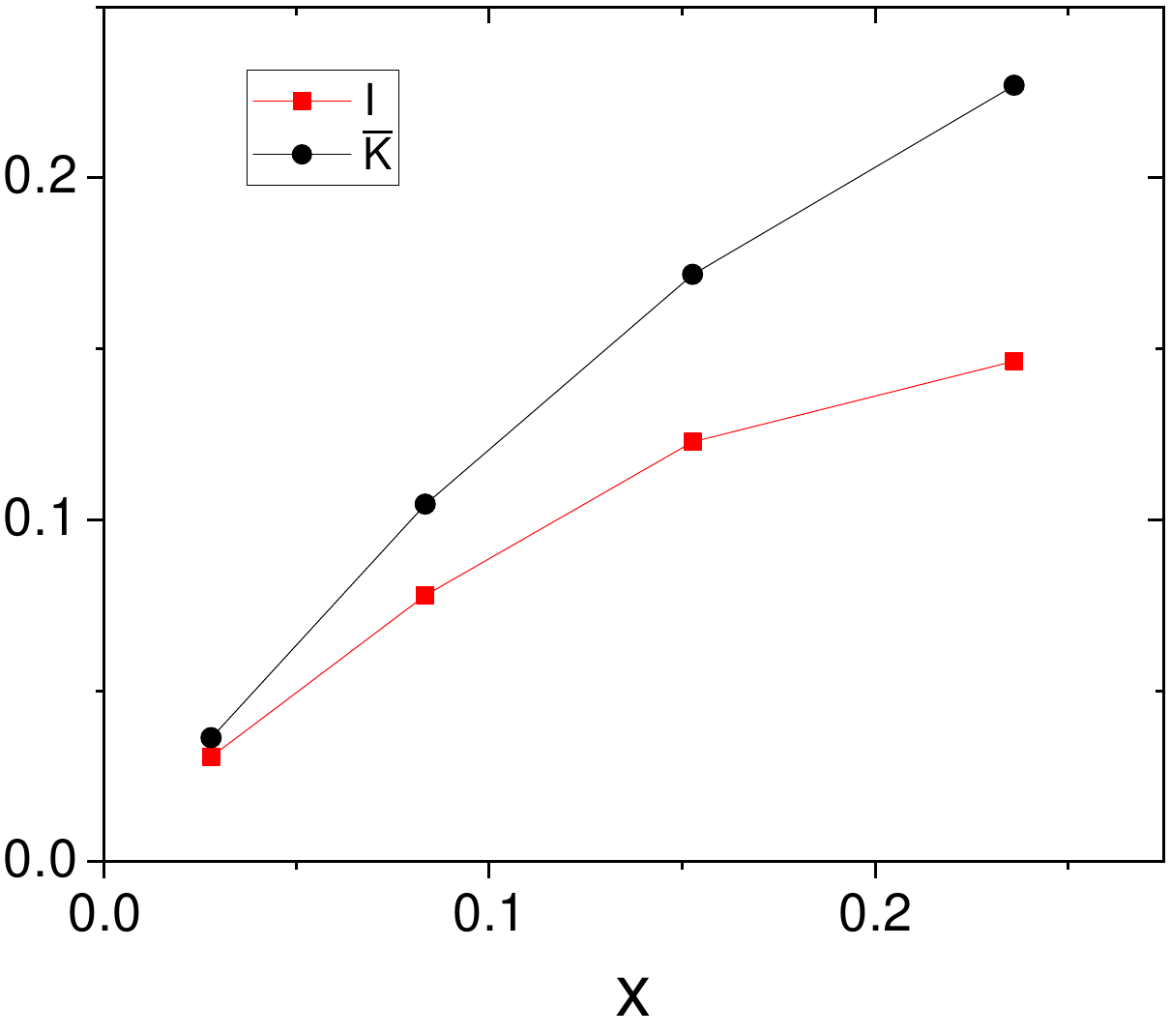}
\caption{Evolution of the total optical spectral weight $\bar{K}$ and the integrated incoherent spectral weight $I$ with hole doping  in the 2D $t-J$ model calculated from the dynamical variational theory. Both $\bar{K}$ and $I$ are measured in unit of $Nt$. Note that $I$ is now always smaller than $\bar{K}$. The calculation is done on a $12\times12$ cluster with periodic boundary condition. Here we have neglected the antiferromagnetic ordering at low doping.}
\end{figure}

In Fig.6, we plot the evolution of the total optical weight $\bar{K}$ and the integrated incoherent optical weight $I$ with hole concentration calculated from the dynamical variational theory. Both $\bar{K}$ and $I$ are found to increase monotonically with doping. It should be emphasized that $I$ is now always smaller than $\bar{K}$. The dynamical variational theory presented here thus indeed solve the inconsistency in the mean field treatment. 

\begin{figure}
\includegraphics[width=7.5cm]{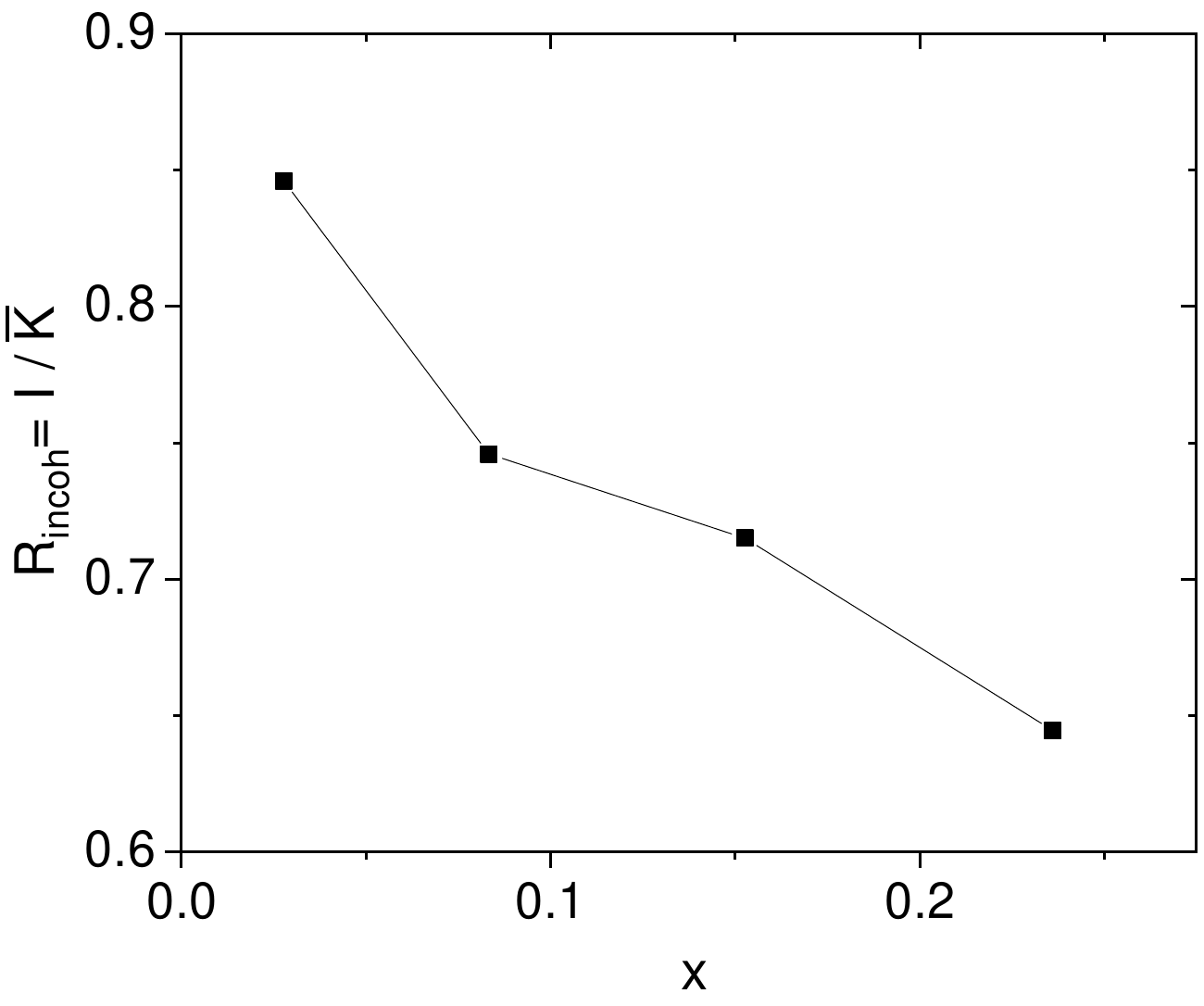}
\caption{Evolution of the ratio of the incoherent spectral weight with hole doping  in the 2D $t-J$ model calculated from the dynamical variational theory. The calculation is done on a $12\times12$ cluster with periodic boundary condition. Here we have neglected the antiferromagnetic ordering at low doping.}
\end{figure}

In Fig.7, we present the evolution of the ratio of the incoherent spectral weight with hole doping. $R_{incoh}$ is found to decrease monotonically with hole doping but remains larger than 60$\%$ for $x>0.2$. Thus, most part of the electron spectral weight remains incoherent throughout the phase diagram. This implies that the electron fractionalization mechanism studied in this paper has already accounted for a major part of the origin of electron incoherence in the 2D $t-J$ model. The evolution of the Drude weight $D=\bar{K}-I$ is plotted as a function of $x$ in Fig.8. Interestingly, we find that $D$ scales almost linearly with $x$.
 
\begin{figure}
\includegraphics[width=7.5cm]{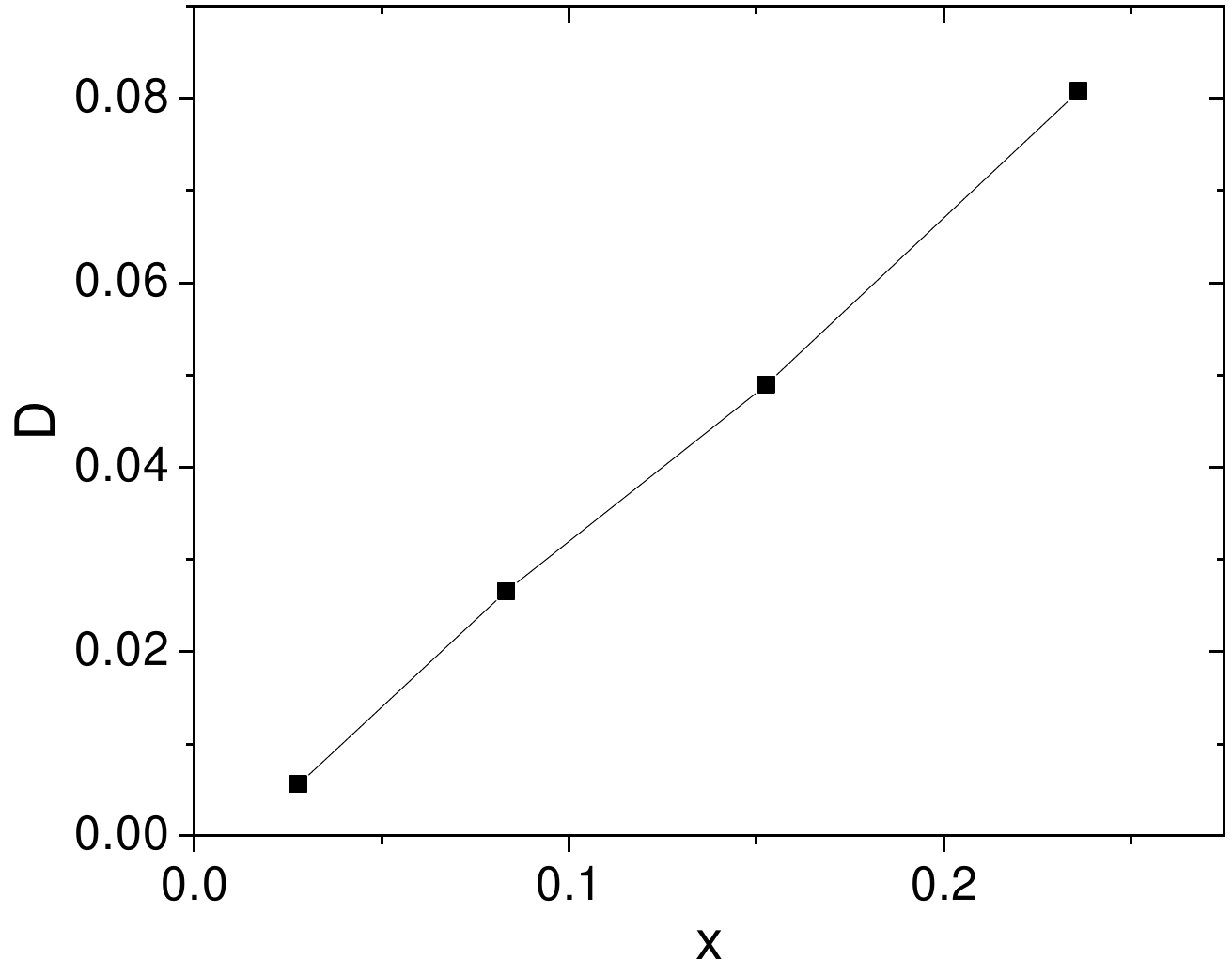}
\caption{Evolution of the Drude weight with hole doping in the 2D $t-J$ model calculated from the dynamical variational theory. The calculation is done on a $12\times12$ cluster with periodic boundary condition. Here we have neglected the antiferromagnetic ordering at low doping.}
\end{figure}

\section{Conclusions and Discussions}
For a translational invariant electron system, the optical absorption at finite frequency is a direct measure of the electron incoherence induced by electron correlation effect. In this work, we have studied the optical conductivity of the 2D $t-J$ model from both the slave Boson mean field theory and a dynamical variational theory. The dynamical variational theory is constructed following the mean field excitation picture and it treats the Gutzwiller projected mean field excitations as the basis function of a variational subspace. We find that the electron fractionalization inherent in such a theoretical framework can generate strong electron incoherence. However, a mean field treatment of such an effect results in the unphysical prediction of a negative Drude weight. We find that such an inconsistency in the slave Boson mean field theory can be cured by working in the variational subspace spanned by the Gutzwiller projected mean field excitations.  

We find that both the total optical weight $\bar{K}$ and the integrated incoherent optical weight $I$ increase monotonically with doping, with their ratio $R_{incoh}=I/\bar{K}$ decreases monotonically with doping. Such a doping evolution is consistent with the general expectation for a doped Mott insulator. We note that most part of the optical spectral weight remains incoherent throughout the phase digram. In particular, $R_{incoh}>0.6$ even for $x>0.2$. We thus believe that the majority part of the electron incoherence in the 2D $t-J$ model can be attributed to the electron fractionalization mechanism. In the slave Boson mean field picture, the electron fractionalization manifests itself in the current excitation process in the form of holon backflow accompanying the excited spinon.

We find that the Drude weight deduced from $D=\bar{K}-I$ scales linearly with $x$ throughout the phase diagram. Since we are working in the superconducting state, $D$ is also the superfluid density $\rho_{s}$. From our calculation, we do not see any trend of a non-monotonic doping dependence in the superfluid density. We feel that such a non-monotonic behavior in $\rho_{s}$ should be attributed to disorder effect. In a related work, we do find that small amount of Zinc impurity can already generate non-monotonic doping dependence in the superfluid density.     

We note that while the dynamical variational theory presented in this paper can provide a satisfactory description on the absolute scale and its doping evolution of integrated incoherent optical weight, the spectral shape of $\sigma^{reg}(\omega)$ is less reliable. In particular, the theory fails to produce the well known "conformal tail" feature in the optical conductivity, namely, the scale invariant form $\sigma^{reg}(\omega)\sim \omega^{-\alpha}$ in the large $\omega$ regime\cite{Marel,Heuman}. We think that the multi-spinon excitation processes that are neglected in our variational treatment may be responsible for the generation of such a "conformal tail" feature. Such multi-spinon excitation processes are expected to play an essential role in the physics of the cuprate superconductors as recent RIXS measurements find that spin-wave-like paramagnon fluctuation persists even in strongly overdoped cuprate systems\cite{Tacon,Dean}, which implies the importance of spinon interaction effect. It is a big theoretical challenge to integrate such strong paramagnon fluctuation into the standard fermionic RVB scheme.

Beside the optical conductivity, the electron incoherence also manifests itself in other observation channels such as the single particle spectrum and density fluctuation spectrum. In particular, recent EELS measurement indicates that the density fluctuation in the cuprate system is qualitatively different from what would expect for a standard Fermi liquid metal\cite{Mitrano}. More specifically, while the density fluctuation in a Fermi liquid metal is restricted to a small energy window of the order of $v_{\mathrm{F}}q$ in the long wave length limit, the observed density fluctuation spectrum is almost featureless and extends to energy as high as $1\ eV$. Qualitatively similar broad spectrum continuum had been reported in early Raman scattering measurements.

In principle, the density fluctuation spectrum of the 2D $t-J$ model can be calculated in the same dynamical variational scheme as presented in this paper. The density fluctuation spectrum of the system is defined by
\begin{equation}
S(\mathbf{q},\omega)=-2\Im D(\mathbf{q},\omega+i0^{+})
\end{equation}
Here $D(\mathbf{q},\omega+i0^{+})$ is the Fourier transform of the retarded green function of the density operator defined as
\begin{equation}
D(\mathbf{q},t)=-i\theta(t)\langle [\rho_{\mathbf{q}}(t), \rho_{-\mathbf{q}}(0)] \rangle
\end{equation}
in which 
\begin{equation}
\rho_{\mathbf{q}}=\sum_{i}n_{i}e^{i\mathbf{q}\cdot\mathbf{r}_{i}}
\end{equation}
In the long wavelength limit, the density fluctuation can be related to the current fluctuation through the continuity equation
\begin{equation}
\frac{\partial\rho}{\partial t}+\nabla\cdot j=0
\end{equation} 
Thus the density fluctuation spectrum constitute an important complementary information on the origin of the electron incoherence in the cuprate systems. However, we note that different from the current operator, the density operator at $\mathbf{q}=0$ is a strictly conserved quantity and can not generate any excitation. For $\mathbf{q}\neq 0$, the density fluctuation is composed of the contribution from both coherent quasiparticle excitations and incoherent electron continuum. The contribution from the coherent quasiparticle is expected to exhibit strong momentum and energy dependence. Currently it is not clear if the observed density fluctuation spectrum is consistent with prediction from the calculation on a translational invariant 2D $t-J$ model.

\begin{acknowledgments}
We acknowledge the support from the National Natural Science Foundation of China(Grant No. 12274453).
\end{acknowledgments}

\end{document}